# First-principles high-throughput screening of shape-memory alloys based on energetic, dynamical, and structural properties

Joohwi Lee[1,*], Yuji Ikeda[2], and Isao Tanaka[1,2,3,4]

[1]*Department of Materials Science and Engineering, Kyoto University, Kyoto, 606-8501, Japan*

[2]*Elements Strategy Initiative for Structure Materials (ESISM), Kyoto University, Kyoto 606-8501, Japan*

[3]*Center for Materials Research by Information Integration, National Institute for Materials Science (NIMS), Tsukuba 305-0047, Japan*

[4]*Nanostructures Research Laboratory, Japan Fine Ceramics Center, Nagoya 456-8587, Japan*

* Corresponding author. Tel.: +81 75 753 5435

E-mail address: lee.joohwi@gmail.com (Joohwi Lee)

**ABSTRACT**

First-principles-based materials screening is systematically performed to discover new combinations of chemical elements possibly making shape-memory alloys (SMAs). The B2, $D0_3$, and $L2_1$ crystal structures are considered as the parent phases, and the 2H and 6M structures are considered as the martensitic phases. 3,384 binary and 3,243 ternary alloys (6,627 in total) with stoichiometric composition ratios are investigated by the materials screening in terms of energetic and dynamical stabilities of the martensitic phases as well as structural compatibility between the parent and the martensitic phases. 187 alloys are found to survive after the screening. Some of the surviving alloys are constituted by the chemical elements already widely used in SMAs, but other various metallic elements are also found in the surviving alloys. The energetic stability of the surviving alloys is further analyzed by comparison with the data in Materials Project Database (MPD) to examine the alloys which may occur phase separation or transition.

## I. INTRODUCTION

Shape-memory alloys (SMAs) constitute an important class of materials in industrial use because of their shape-memory effects and pseudoelasticity [1]. Already various kinds of SMAs have been well known. Ni–Ti (Nitinol) alloys are now widely used because of its working range around room temperature, good pseudoelastic property, phase stability, and so on [2]. The Ni–Ti alloys, however, suffer from large thermal hysteresis unless additional elements are included [3]. Toxicity of Ni in these alloys is also problematic for bioapplications [4]. Cu-based SMAs such as Cu–Zn, Cu–Al, and their multicomponent alloys are commercially available because of their low price, but they also have disadvantages such as instability of martensitic phase, brittleness [5], and poor thermomechanical performance [6]. Discovery of new SMAs are therefore still needed for better functional stability, design of working temperature [7,8], and other special purposes such as nontoxic biomaterials [4] and ferromagnetic SMAs [9].

To design SMAs, their working temperature and functional stability may be the most important engineering properties. Working temperature is critical especially to design high temperature SMAs (HTSMAs). Functional stability is also important to improve the reliability of SMAs. We can expect the working temperature is related to the energetic stability of the martensitic phase relative to that of the parent phase (see Sec. III.A). Meanwhile, some reports have found that better structural compatibility between the parent and the martensitic phases results in smaller thermal hysteresis, which gives better functional stability [10,11].

We can adjust working temperature and functional stability of SMAs by replacing their constituent elements with others. The working temperature of Ni–Ti alloys can be much increased to the range of 400–1200 K by the total or partial replacement of Ni and Ti with the same group elements, namely Pd or Pt [12,13] and Zr or Hf [14], respectively. These reports imply that the martensitic phases become energetically more stable than the parent phases at the low temperature by replacing constituent elements. Meanwhile, better functional stability, which is associated with smaller thermal hysteresis and functional fatigue, can be achieved by adding Cu [11] or Pd [10] in Ni–Ti alloys and Cr in Ti–Pd alloys [15].

To discover new materials by modifying their constituent elements, computational research is useful because of its efficiency compared with experimental search. Especially, materials screening based on systematic first-principles calculations prior to experimental synthesis is useful to limit the searching space.

Thanks to the recently improved computational machine power, now we can perform high-throughput first-principles calculations for various combinations of chemical elements with various crystal structures for searching new materials. Actually many different materials systems have been investigated in this manner [16-19]. Hautier *et al.* [16] and Hinuma *et al.* [17] have identified new ternary-oxide and zinc-nitride semiconductors, respectively, using the "prototype" crystal structures of Inorganic Crystal Structure Database (ICSD) [20]. Carrete *et al.* [18] have reported the semiconductors with the half-Heusler structure that show low thermal conductivity. Greeley *et al.* [19] have reported the binary surface alloys to show a good electrocatalytic property.

In this study, we perform the first-principles high-throughput screening to discover new combinations of chemical elements that are possibly available as SMAs. As the screening conditions, we consider the energetic and dynamical stabilities of the martensitic phase, as well as the structural compatibility between the parent and the martensitic phases, which is probably related to the functional stability. These screening conditions require relatively low computational costs and are therefore suitable for the high-throughput materials screening. 6,627 alloys are initially considered, and 187 survive as the candidates of new SMAs after the screening. We investigate which elements mainly participate in the surviving alloys. Finally, the energetic stability of surviving candidates are further analyzed by comparison with the first-principles data in Materials Project Database (MPD) [21] to examine the alloys which may occur phase separation or transition.

## II. COMPUTATIONAL DETAILS

### A. Elements and crystal structures of investigated alloys

We consider binary and ternary alloys with stoichiometric composition ratios composed of 48 metallic elements between Li and Bi. Specifically, we consider 1,128 (= ${}_{48}C_2$) $XY$, 2,256 (= ${}_{48}P_2$) $X_3Y$, and 3,243 (= 3 × ${}_{47}C_2$) $X_2YZ$ ($X$ = Ti, Cu, Zn) alloys with the B2, D0$_3$, and L2$_1$ parent-phase structures, respectively. These parent-phase structures are derived from the body-centered cubic (bcc) structures [22] as described in Fig. 1. For the martensitic phases, we consider the orthorhombic 2H (or B19 for the binary alloys with the B2 parent phase) and monoclinic 6M [in the Otsuka notation [23,24], which will be used hereafter to correctively refer to 9R (for B2) and 18R (for D0$_3$ and L2$_1$) in the Ramsdell notation [25]] structures. Both the 2H and 6M martensitic-phase structures have the close-packed basal plane, but they have different stacking orders: "AB" for the 2H structure and "ABCBCACAB" for the 6M structure. Figure 2(a) shows the crystal structures of the 2H and 6M for ternary $X_2YZ$ alloys with the $L2_1$ parent phase. The structure of the 2H is explicitly calculated in this study, while the structure of the 6M is estimated from that of the 2H as described later. The space-group type of the 2H structure is *Pmma* for the B2-parent *XY* alloys and *Pnma* for the $D0_3$-parent $X_3Y$ and for the $L2_1$-parent $X_2YZ$ alloys.

The lattice basis for the 2H martensitic structure is given as

$$\mathbf{L}^{2H} = (\mathbf{a}^{2H}\,\mathbf{b}^{2H}\,\mathbf{c}^{2H}) = \begin{pmatrix} a^{2H} & 0 & 0 \\ 0 & b^{2H} & 0 \\ 0 & 0 & c^{2H} \end{pmatrix}, \tag{1}$$

where $a^{2H}$, $b^{2H}$, and $c^{2H}$ are the lattice constants of the 2H structure. The unit cell of the parent phase that changes to $\mathbf{L}^{2H}$ after the martensitic transformation may be given as

$$\mathbf{L}^{p\to 2H} = \begin{pmatrix} \mathbf{a}^p - \mathbf{b}^p & -\mathbf{c}^p & \mathbf{a}^p + \mathbf{b}^p \end{pmatrix} = \begin{pmatrix} a^p & 0 & a^p \\ -a^p & 0 & a^p \\ 0 & -a^p & 0 \end{pmatrix} \quad \text{(for B2)} \tag{2}$$

and

$$\mathbf{L}^{p\to 2H} = \begin{pmatrix} (\mathbf{a}^p - \mathbf{b}^p)/2 & -\mathbf{c}^p & (\mathbf{a}^p + \mathbf{b}^p)/2 \end{pmatrix} = \begin{pmatrix} a^p/2 & 0 & a^p/2 \\ -a^p/2 & 0 & a^p/2 \\ 0 & -a^p & 0 \end{pmatrix} \quad \text{(for D0}_3\text{ and L2}_1\text{)}, \tag{3}$$

where $\mathbf{a}^p$, $\mathbf{b}^p$, and $\mathbf{c}^p$ are the lattice basis of the conventional unit cell for the parent phase structures, and $a^p$ is their lattice constant. The deformation gradient [26] $\mathbf{F}^{2H}$ is then obtained as

$$\mathbf{F}^{2H} = \mathbf{L}^{2H}(\mathbf{L}^{p\to 2H})^{-1}. \tag{4}$$

Figure 2(b) describes the martensitic transformation between the $L2_1$ parent and its 2H martensitic phases. It should be emphasized that the martensitic transformation path is similar to the Burgers path [27,28] for pure metals, which describes the transformation between the bcc and the hexagonal close-packed (hcp) structures.

In order to reduce computational costs, the crystal structure of the 6M is estimated from that of the 2H as follows. We first assume that their basal-plane structure and layer distance along the stacking direction are the same for both the 2H and 6M. We further assume that the stacking position of each layer is different exactly by $a^{2H}/3$ along $\mathbf{a}^{2H}$. Then, the lattice basis for the 6M structure is given as

$$\mathbf{L}^{6M} = (\mathbf{a}^{6M}\ \mathbf{b}^{6M}\ \mathbf{c}^{6M}) = \begin{pmatrix} a^{2H} & 0 & -a^{2H}/3 \\ 0 & b^{2H} & 0 \\ 0 & 0 & 3c^{2H} \end{pmatrix}. \quad (5)$$

The unit cell of the parent phase that changes to the $\mathbf{L}^{6M}$ after the martensitic transformation may be given as

$$\mathbf{L}^{p\to 6M} = \begin{pmatrix} \mathbf{a}^p - \mathbf{b}^p & -\mathbf{c}^p & 3(\mathbf{a}^p + \mathbf{b}^p) \end{pmatrix} = \begin{pmatrix} a^p & 0 & 3a^p \\ -a^p & 0 & 3a^p \\ 0 & -a^p & 0 \end{pmatrix} \quad \text{(for B2)} \quad (6)$$

and

$$\mathbf{L}^{p\to 6M} = \begin{pmatrix} (\mathbf{a}^p - \mathbf{b}^p)/2 & \mathbf{c}^p & 3(\mathbf{a}^p + \mathbf{b}^p)/2 \end{pmatrix} = \begin{pmatrix} a^p/2 & 0 & 3a^p/2 \\ -a^p/2 & 0 & 3a^p/2 \\ 0 & -a^p & 0 \end{pmatrix} \quad \text{(for D0}_3\text{ and L2}_1\text{)}. \quad (7)$$

The deformation gradient $\mathbf{F}^{6M}$ is then obtained as

$$\mathbf{F}^{6M} = \mathbf{L}^{6M}(\mathbf{L}^{p\to 6M})^{-1}. \quad (8)$$

### B. Materials-screening conditions

Figure 3 shows the flowchart of the materials screening.

Firstly, we check whether the space-group type of the optimized structure of the 2H martensitic phase ($\mathbf{G}^{m}{}_{\rm opt}$) is actually the same as the expected one ($\mathbf{G}^{p}{}_{\rm init}$). For many investigated alloys, the 2H structure is optimized to the parent-phase structure or to some other structure. Such alloys are excluded from the screening because they cannot have the assumed martensitic structure. The space-group types of the optimized structures are checked using the SPGLIB library inside the PHONOPY code [29,30].

Secondly, we investigate the energetic stability. Here we check whether the energy of the martensitic phase ($E^m$) is smaller than that of the parent phase ($E^p$). Only the alloys satisfying $\Delta E^{m\text{-}p} \equiv E^m - E^p < 0$ survive. We also guarantee that the 2H martensitic phase is energetically more stable than pure metals as references. Here, the formation energy of the martensitic phase relative to those of pure metals in their most stable crystal structures, $E_f^m$, must be smaller than zero.

Thirdly, the structural compatibility between the parent and the martensitic phases is considered. For this purpose, we use the transformation stretch tensors [26] $\mathbf{U}^{\rm 2H}$ and $\mathbf{U}^{\rm 6M}$, which are positive-definite and symmetric matrices. These matrices are obtained from $\mathbf{F}^{\rm 2H}$ and $\mathbf{F}^{\rm 6M}$ using the polar decomposition as

$$\mathbf{F}^{\rm 2H} = \mathbf{R}^{\rm 2H}\mathbf{U}^{\rm 2H} \tag{9}$$

and

$$\mathbf{F}^{\rm 6M} = \mathbf{R}^{\rm 6M}\mathbf{U}^{\rm 6M}, \tag{10}$$

where $\mathbf{R}^{\rm 2H}$ and $\mathbf{R}^{\rm 6M}$ are rotation matrices. James *et al.* have shown using their model that when the second largest eigenvalue $\lambda_2$ of $\mathbf{U}$ (hereafter the $\mathbf{U}^{\rm 2H}$ or $\mathbf{U}^{\rm 6M}$ are collectively referred to as $\mathbf{U}$) is equal to one, the two phases can make a distortionless interface [26], which is intuitively expected as an advantage for showing better functional stability. Actually, several SMAs with small thermal hysteresis and functional fatigue are found by modifying the composition ratios to realize $\lambda_2$ close to one [10,15,31,32]. Based on these reports, we adopt $|\lambda_2 - 1| < 0.01$ as a screening condition. We also consider the volume difference between the parent and the martensitic phases, because the large volume difference is expected to cause huge stress between the two phases and to result in low functional stability [32]. Actually, several materials systems such as lithium-ion batteries, whose applications are related to their phase transitions, are known to have good advantage of reliability when they show small volume differences [33]. The relative difference between the volume of the parent ($V^p$) and the martensitic ($V^m$) phases are obtained as det(**U**), and hence we adopt |det(**U**) − 1| < 0.01 as another screening condition.

Lastly, the dynamical stability of the martensitic phase is investigated. For this purpose, we analyze the phonon frequencies of the martensitic phases $\omega_{ph}{}^{m}$. The alloys with imaginary phonon frequencies, *i.e.*, $\{\omega_{ph}{}^{m}\}^2 < 0$ for some phonon modes, are screened out, because the existence of the imaginary phonon frequencies indicates that the crystal structure is dynamically unstable. The phonon calculations are performed under the harmonic approximation on the lattice Hamiltonian. Force constants of the alloys are calculated from their supercell models based on density functional perturbation theory [34] at the Γ point, and then phonon frequencies are calculated from the force constants. Phonon dispersion curves and density of states (PhDOSs) are used to confirm the dynamical stability of martensitic phase by the existence of the imaginary phonon frequencies. Note that the dynamical stability of the 6M martensitic structure is assumed to be the same as that of the 2H martensitic structure. The phonon calculations are performed using the PHONOPY code [29,30].

### C. First-principles calculations

The first-principles calculations are performed by the project augmented wave (PAW) method [35,36] implemented in the Vienna *Ab-initio* Simulation Package (VASP) [37,38] within the framework of the generalized gradient approximation of Perdew-Burke-Ernzerhof form [39]. The cutoff energy is set to 400 eV. The volume and shape of the cells and internal atomic coordinates are fully relaxed until residual forces acting on atoms reach below 0.005 eV/Å. The structure optimization is performed for a primitive-cell model to reduce computational costs. Table I shows detailed computational conditions for the primitive cell, supercell for the phonon calculations, and ***k***-space sampling. Both the nonmagnetic (NM) and the ferromagnetic (FM) states are calculated for each system, and the lower-energy states are investigated in the subsequent materials screening.

## III. RESULTS AND DISCUSSION

### A. SMAs reported in experiments

Prior to the materials screening, we first investigate $\Delta E^{m\text{-}p}$ for 13 alloys that were reported to show the shape-memory effects near the stoichiometric composition ratios in experiments. Table II summarizes the experimental transformation temperature $T_c$ and the computed $\Delta E^{m\text{-}p}$ for the 13 SMAs, and Fig. 4 shows their correlation. Here $T_c$ is determined as the average over the reported characteristic martensitic transformation temperatures $A_s$, $A_f$, $M_s$, and $M_f$ in experiments [9,12,13,31,40-45]. When some SMAs were reported to have different martensitic structures from 2H in experiments, the energies of these crystal structures are additionally calculated to obtain $\Delta E^{m\text{-}p}$.

We can derive three results from the obtained data. Firstly, the martensitic phases have lower energies than the parent phase for all the 13 investigated SMAs. Since the first-principles calculations give the energies at 0 K, this result supports the experimental fact that the martensitic phases are found at lower temperature than the parent phase for the 13 SMAs. Secondly, most of the energy differences among various martensitic phases are less than 0.01 eV/atom, which are much smaller than the energy differences between the parent and the martensitic phases. This guarantees that the energy of the 2H structure can be used as the representative value among the various martensitic phases for the screening procedure using $\Delta E^{m\text{-}p}$. Thirdly, the computed $\Delta E^{m\text{-}p}$ shows strong correlation with the experimental $T_c$. Specifically, the SMAs with large $|\Delta E^{m\text{-}p}|$ show high $T_c$. The linear correlation coefficient between $\Delta E^{m\text{-}p}$ and $T_c$ is −0.92. This implies that $\Delta E^{m\text{-}p}$ can be used to roughly but efficiently estimate $T_c$ for various kinds of alloys. For example, the $\Delta E^{m\text{-}p}$ of −0.05 eV/atom corresponds to $T_c$ of 400 K.

Figures 5 and 6 show the phonon dispersion curves of parent and martensitic phases, respectively, for the 13 SMAs. For the martensitic phases, we investigate energetically the most stable crystal structure for each of the 13 SMAs. These martensitic structures do not show any imaginary modes, indicating that they are dynamically stable. This satisfies our screening condition that the martensitic structures should be dynamically stable. In contrast, the parent phases of the most of the 13 SMAs show imaginary frequencies in a wide range of wave vectors. This indicates that they are dynamically unstable under the harmonic approximation. The anharmonic phonon effects [46,47], which is significant particularly at high temperature, is probably essential to make the parent phases dynamically stable.

According to these results, we expect that the screening conditions for the energetic and dynamical stabilities described in Sec. II are suitable for the materials selection of SMAs. Later we will also discuss the structural compatibility of the 13 investigated alloys.

### B. Candidates for new SMAs

In order to identify candidates for new SMAs, we apply the screening conditions described in Sec. II.B and Fig. 3 to 6,627 alloys. Figure 7 shows the correlations among four quantities, $\Delta E^{m\text{-}p}$, $E_f^m$, det(**U**) − 1, and $\lambda_2 - 1$, for 3,997 alloys in the set-1. Gaussian kernel density estimation [48] is used to put colors in the plots. There are no strong correlations among the four quantities. The maximum magnitude of the linear correlation coefficient is only 0.42, found between $\Delta E^{m\text{-}p}$ and det(**U**) − 1. This indicates that these four screening conditions work almost independently. The screening conditions are also indicated in Fig. 7. Although our screening conditions are rather tight especially for det(**U**) – 1 and $\lambda_2 - 1$, as many as 187 alloys survive because of the high density of alloys within the screening-condition slot. The lists of surviving alloys with the 2H and 6M martensitic structures are given in Tables III and IV, respectively. Among 187 alloys, 54 and 133 alloys correspond to the 2H and 6M martensitic structures, respectively. The alloys with the 6M martensitic structure more frequently survives.

It is interesting that the 13 SMAs analyzed in Sec. III.A are not included in the list of surviving alloys. We confirm that they are included in the set-3, satisfying the screening condition for the energetic and dynamical stability of the martensitic phase. However, they do not satisfy the remaining two screening conditions related to the structural compatibility ($|\lambda_2 - 1| < 0.01$ and $|\det(\mathbf{U}) - 1| < 0.01$). Their low structural compatibility indicates their low functional reliability. Actually, experimental reports (Table II) show that the thermal hysteresis of these alloys, except for $Zn_2AuCu$, is larger than 20–30 K, which is not small [3,49]. The larger thermal hysteresis should result in the worse functional stability. Inclusion of point defects in off-stoichiometric composition ratios [31] or solute elements may be essential for these alloys for better functional stability or smaller thermal hysteresis. [10,11,15].

Frequency of chemical elements in the surviving 111 binary alloys is summarized in Fig. 8. The ternary $X_2YZ$ alloys are excluded for this analysis because the $X$ component for the ternary in this study is restricted to three elements, *i.e.*, Ti, Cu, and Zn. Among the 48 chemical elements, Cu, Zn, Ag, Au, and Pd are the most frequently included elements in the descending order. The result is natural since many Cu-based alloys are known to exhibit the shape-memory effects. Cu–Zn-based SMAs are found in the form of either binary alloys of 60–64 at.% Cu[50] or ternary alloys incorporating Al, Si, Ga, and Mn [5]. Cu–Al-based SMAs are found in the form of ternary alloys incorporating Ni [51], Mn [52], Be [53], and Zn [5]. Cu–Sn-based binary alloys also show the shape-memory effects for 74–91 at.% Cu [5]. These alloys are not included in the 13 SMAs analyzed in Sec III.A. This may be because their shape-memory effects were reported experimentally only for the off-stoichiometric composition ratios.

Besides these popular alloy systems, many other elements and their combinations are found in the surviving alloys. Among such elements, Li and Sc have been rarely used as the constituent elements of the SMAs reported in experiments. Only recently, the $Mg_{80}Sc_{20}$ alloy was discovered as a SMA having technological advantages with its light weight [54]. The In–Tl nanowire was also reported to exhibit the shape-memory effects [55]. Much opportunity to discover new SMA from the less popular systems can be expected.

According to several review papers [7,8], the SMAs with the $T_c$ above 370–400 K are categorized in HTSMAs. Since the $\Delta E^{m\text{-}p}$ of −0.05 eV/atom approximately corresponds to the $T_c$ of 400 K as described in Sec III.A, it might be interesting to classify the surviving alloys at $\Delta E^{m\text{-}p} = -0.05$ eV/atom. Among 111 binary alloys, 60 alloys show $\Delta E^{m\text{-}p} < -0.05$ eV/atom. Their constituent elements are separately shown in Fig. 8. Al, Sc, Ti, Zn, Hf, and Pt are found five or more times in the surviving alloys with $\Delta E^{m\text{-}p} < -0.05$ eV/atom. These chemical elements, except for Zn, are not so much included in the chemical combinations with $0 > \Delta E^{m\text{-}p} > -0.05$ eV/atom. This indicates that these elements are preferable to form HTSMAs. As a matter of fact, Pt-rich alloys such as Pt–Al and Pt–Ga [56] are used for HTSMAs.

Finally, energetic stability of the surviving 187 compounds is examined. The stability is already examined by the formation energy $E_f^m$ with respect to the pure metals as described in Sec. II.B. Besides $E_f^m$, the energy of the alloys may be compared with that on the convex-hull energy for the corresponding binary and ternary systems. Since the construction of the convex-hull energy for many systems is computationally laborious, we refer to the data of the MPD [21], which were obtained under similar computational conditions with the present study. The convex-hull energy of the corresponding chemical composition with reference to pure metals in MPD, $E_f^{\mathrm{conv}}$, is compared to the present $E_f^m$. Tables III and IV show the difference $E_f^m - E_f^{\mathrm{conv}}$, together with the corresponding phases for the given composition. Since most of the 2H and 6M structures for the surviving 187 compounds are missing in MPD, the structure on the convex hull in MPD and that of the present martensitic phase are mostly different. The small absolute value of $E_f^m - E_f^{\mathrm{conv}}$ of less than 0.1 eV/atom may be ascribed to either the difference in the structure or the detailed computational conditions. However, if the $E_f^m - E_f^{\mathrm{conv}}$ exceeds approximately 0.1 eV/atom, the alloy may be difficult to be formed or susceptible to the phase separation or transition.

## IV. CONCLUSION

We perform the first-principles-based materials screening to discover so far unknown combinations of chemical elements possibly making SMAs. As the screening conditions, we consider the energetic and dynamical stabilities of the crystal structure in the martensitic phase as well as the structural compatibility of the crystal structures between the parent and the martensitic phases.. The B2, $D0_3$, and $L2_1$ crystal structures are considered as the parent phase, while the 2H and the 6M structures are considered as the martensitic phase.

6,627 alloys with binary and ternary combinations of chemical elements are screened, resulting in 187 candidates. The surviving candidates are composed of various chemical combinations, indicating that the elements in the wide range in the periodic table should be investigated to discover new SMAs. The formation energies of the surviving candidates are further compared with those of first-principles data in MPD to prove out the alloys which may occur phase separation or transition.

We also examine the correlation between the martensitic transformation temperature $T_c$ in experiments and the $\Delta E^{m\text{-}p}$ obtained from the first-principles calculations for the 13 SMAs with nearly stoichiometric composition ratios. Strong correlation is found between the experimental $T_c$ and the computed $\Delta E^{m\text{-}p}$. This implies that $\Delta E^{m\text{-}p}$ can be used to roughly estimate the working temperature range as SMAs.

The findings in this study may help the new discovery of SMAs that overcome the problems of costs, toxicity, and poor functional stability. Furthermore, this study shows the strategy for the design of SMAs based on first-principles calculations. Although in this study we focus on only the alloys with stoichiometric composition ratios, the strategy in this study may be able to be applied also to the alloys with nonstoichiometric composition ratios. Actually, the off-stoichiometric $Zn_{45}Au_{30}Cu_{25}$ alloy shows smaller thermal hysteresis and functional fatigue than the stoichiometric $Zn_2AuCu$ alloy [31], and hence such composition dependence must be also very interesting.

## ACKNOWLEDGEMENTS

This work was supported by Grant-in-Aid for Scientific Research (A) and Grant-in-Aid for Scientific Research on Innovative Areas "Nano Informatics" (Grant No. 25106005) from the Japan Society for the Promotion of Science (JSPS), and Support program for starting up innovation hub on Materials research by Information Integration" Initiative from Japan Science and Technology Agency. JL acknowledges Grant-in-Aid for International Research Fellow of JSPS (Grant No. 2604376) and JSPS fellowships. YI acknowledges Grant-in-Aid for Young Scientist (B) of JSPS (Grant No. 16K18228). Funding by the Ministry of Education, Culture, Sports, Science and Technology (MEXT), Japan, through Elements Strategy Initiative for Structural Materials (ESISM) of Kyoto University, is also gratefully acknowledged.

Table I. Detailed information on the investigated structures as well as on the computational conditions. *N* denotes the number of atoms in the primitive cell.

| Phase | Index | Binary *XY* (B2 parent) | Binary $X_3Y$ ($D0_3$ parent) | Ternary $X_2YZ$ ($L2_1$ parent) |
|---|---|---|---|---|
| Parent phase | Space group | *Pm*-3*m* | *Fm*-3*m* | *Fm*-3*m* |
| | *N* | *X*: 1, *Y*: 1 | *X*: 3, *Y*: 1 | *X*: 2, *Y*: 1, *Z*: 1 |
| | ***k***-space sampling | 16×16×16, Γ-centered | 16×16×16, Γ-centered | 16×16×16, Γ-centered |
| 2H martensitic phase | Space group | *Pmma* | *Pnma* | *Pnma* |
| | *N* | *X*: 2, *Y*: 2 | *X*: 6, *Y*: 2 | *X*: 4, *Y*: 2, *Z*: 2 |
| | ***k***-space sampling | 12×12×12, Γ-centered | 12×12×12, Γ-centered | 12×12×12, Γ-centered |
| | Supercell size for phonon calculations | 2×4×2 of primitive cell | 2×2×2 of primitive cell | 2×2×2 of primitive cell |

Table II. Characteristic martensitic transformation temperatures for 13 experimentally reported SMAs with nearly stoichiometric composition ratios. The data correspond to those in Fig. 4.

| This study | | Experimental reports | | | | | | |
|---|---|---|---|---|---|---|---|---|
| Calculated composition | $\Delta E^{m\text{-}p}$ (eV/atom) | Experimental composition | $M_s$ (K) | $M_f$ (K) | $A_s$ (K) | $A_f$ (K) | $T_c$ (K) | Ref. |
| AgCd [a] | −0.005 | $Ag_{51}Cd_{49}$ | | | | | ~223 | [41] |
| AuCd | −0.005 | $Au_{50}Cd_{50}$ | ~293 | ~323 | | | ~308 | [40] |
| TiAu | −0.078 | $Ti_{50}Au_{50}$ | 834 | 823 | 852 | 861 | 843 | [12] |
| $Zn_2AuCu$ | −0.012 | $Zn_{45}Au_{30}Cu_{25}$ | 235 | 233 | 235 | 237 | 235 | [31] |
| $Co_2NiGa$ | −0.042 | $Co_{50}Ni_{24}Ga_{26}$ | ~323 | ~348 | | | ~336 | [42] |
| $Ti_3Nb$ | −0.024 | $Ti_{76}Nb_{24}$ | ~338 | | | | ~338 | [43] |
| $Ni_2MnGa$ | −0.006 | $Ni_{50}Mn_{25}Ga_{25}$ | 275 | | 281 | | 278 | [9] |
| TiNi | −0.043 | $Ti_{50}Ni_{50}$ | 218 | 240 | 269 | 280 | 252 | [12] |
| TiPd | −0.082 | $Ti_{50}Pd_{50}$ | 783 | 717 | 757 | 787 | 761 | [12] |
| TiPt | −0.155 | $Ti_{50}Pt_{50}$ | 1307 | 1257 | 1277 | 1322 | 1291 | [12] |
| $Ti_2NiPd$ | −0.067 | $Ti_{50.5}Ni_{24.5}Pd_{25}$ | 452 | 440 | 457 | 468 | 454 | [13] |
| $Ti_2NiPt$ | −0.087 | $Ti_{50.5}Ni_{24.5}Pt_{25}$ | | 663 | | 778 | 721 | [44] |
| $Ti_2PtIr$ | −0.120 | $Ti_{50}Pt_{25}Ir_{25}$ | 1383 | 1341 | 1394 | 1463 | 1395 | [45] |

[a] Detailed characteristic temperatures are not reported in ref. [41].

Table III. Surviving 54 alloys with the 2H martensitic structure after the materials screening.

| | This study | | | Data from MPD [21] | Difference |
|---|---|---|---|---|---|
| | Alloy | $\Delta E^{m\text{-}p}$ (eV/atom) | $E_f^m$ (eV/atom) | Reported convex hull [a] | $E_f^m - E_f^{conv}$ (eV/atom) |
| *XY* | NaTl | −0.013 | −0.126 | NaTl (*Fd*-3*m*) | −0.010 |
| $X_3Y$ | $Tl_3Bi$ | 0.000 | −0.029 | $Tl_3Bi$ ($P6_3/mmc$) | −0.002 |
| | $Tl_3Na$ | −0.018 | −0.088 | $Tl_3Na$ (*Pm*-3*m*) | −0.005 |
| | $Tl_3Ca$ | −0.018 | −0.224 | $Tl_3Ca$ (*Pm*-3*m*) | 0.019 |
| | $Au_3In$ | −0.038 | −0.136 | $Au_3In$ (*Fm*-3*m*) | −0.038 |
| | $Mg_3Zn$ | −0.041 | −0.037 | Mg + $Mg_{21}Zn_{25}$ | 0.020 |
| | $Zn_3Cu$ | −0.053 | −0.075 | $Zn_8Cu_5$ + Zn | −0.010 |
| | $Cd_3Ag$ | −0.053 | −0.045 | $Cd_8Ag_5$+Cd | −0.004 |
| | $Cd_3Pt$ | −0.054 | −0.199 | CdPt + Cd | −0.037 |
| | $Cu_3Sn$ | −0.057 | −0.002 | Cu + CuSn | 0.012 |
| | $Zn_3Ag$ | −0.058 | −0.045 | $Zn_8Ag_5$ + Zn | −0.014 |
| | $Zn_3Co$ | −0.059 | −0.041 | $Zn_{13}Co$ + Co | −0.001 |
| | $Cd_3Rh$ | −0.061 | −0.087 | Cd + Rh | −0.087 |
| | $Zn_3Au$ | −0.070 | −0.159 | $Zn_3Au$ (*Pm*-3*n*) | −0.005 |
| | $Be_3Pt$ | −0.080 | −0.477 | $Be_{12}Pt$ + Pt | −0.393 |
| | $Zn_3Rh$ | −0.082 | −0.313 | $Zn_3Rh$ ($P6_3/mmc$) | 0.016 |
| | $Sc_3In$ | −0.093 | −0.258 | $Sc_3In$ ($P6_3/mmc$) | 0.047 |
| | $Sc_3Al$ | −0.094 | −0.247 | $Sc_3Al$ (*Pm*-3*m*) | 0.022 |
| | $Pd_3Y$ | −0.096 | −0.796 | $Pd_3Y$ (*Pm*-3*m*) | 0.070 |
| | $Zn_3Ir$ | −0.099 | −0.219 | $Zn_3Ir$ (*I*4/*mmm*) | −0.004 |
| | $Zn_3Ru$ | −0.100 | −0.088 | $Zn_3Ru$ (*I*4/*mmm*) | 0.051 |
| | $Sc_3Tl$ | −0.120 | −0.187 | Sc + Tl | −0.187 |
| | $Ru_3Sc$ | −0.130 | −0.086 | Ru + $ScRu_2$ | 0.210 |
| | $Sc_3Zr$ | −0.136 | −0.007 | Sc+Zr | −0.007 |
| | $Sc_3Sn$ | −0.151 | −0.436 | Sc + $Sc_5Sn_3$ | 0.020 |
| | $Sc_3Pb$ | −0.163 | −0.271 | Sc + $Sc_5Pb_3$ | 0.045 |
| | $Rh_3Y$ | −0.181 | −0.405 | Rh + $Rh_2Y$ | 0.151 |
| | $Hf_3Sc$ | −0.208 | −0.004 | Sc + $Hf_5Sc$ | 0.008 |
| | $Os_3Ti$ | −0.333 | −0.136 | TiOs + Os | 0.221 |
| | $Os_3Hf$ | −0.352 | −0.108 | HfOs + Os | 0.247 |
| | $Os_3Zr$ | −0.370 | −0.040 | $ZrOs_2$ + Os | 0.254 |
| $Zn_2YZ$ | $Zn_2ScCr$ | −0.003 | −0.088 | $Cr + ScZn_2$ | 0.082 |
| | $Zn_2CrZr$ | −0.007 | −0.022 | $Cr + ZrZn_2$ | 0.125 |
| | $Zn_2LiGa$ | −0.029 | −0.176 | $LiZn_3 + LiGa_3$ | −0.009 |
| | $Zn_2BeNi$ | −0.053 | −0.126 | $BeNi + Be_3Ni + Zn_{11}Ni_2$ | 0.114 |
| | $Zn_2MgAg$ | −0.057 | −0.127 | $Mg_2Zn_{11} + MgZn_2 + MgZnAg_2$ | 0.003 |
| | $Zn_2AlNi$ | −0.142 | −0.246 | $Zn + Zn_{11}Ni_2 + Al_4Ni_3$ | 0.085 |
| | $Zn_2AlCo$ | −0.164 | −0.160 | Zn + AlCo | 0.082 |
| | $Zn_2CoGa$ | −0.186 | −0.080 | $Zn + GaCo + Ga_3Co$ | 0.060 |
| $Cu_2YZ$ | $Cu_2InAu$ | −0.017 | −0.073 | $Cu + Cu_3Au + In_2Au$ | 0.037 |
| | $Cu_2ZnGa$ | −0.021 | −0.101 | $Zn_8Cu_5 + GaCu_3 + Ga_2Cu$ | −0.004 |
| | $Cu_2AlAg$ | −0.026 | −0.097 | $Al + AlCu + Al_4Cu_9$ | 0.062 |
| | $Cu_2GaAg$ | −0.028 | −0.036 | $Ag + GaCu_3 + GaAg_2$ | 0.041 |
| | $Cu_2LiSb$ | −0.040 | −0.189 | $Cu + Sb + Li_2Sb$ | 0.024 |
| | $Cu_2GaAu$ | −0.054 | −0.140 | $GaCu_3 + Cu_3Au + GaAu$ | −0.020 |
| | $Cu_2PdSn$ | −0.064 | −0.203 | $Cu + SnPd + Sn_2Pd$ | 0.066 |
| | $Cu_2NiSn$ | −0.093 | −0.061 | $Cu + Ni_3Sn_2 + Ni_3Sn_4$ | 0.058 |
| | $Cu_2SnPt$ | −0.096 | −0.198 | $Cu + SnPt + Sn_3Pt_2$ | 0.095 |
| | $Cu_2AlRu$ | −0.105 | −0.123 | $Cu + AlRu + Al_2Ru$ | 0.224 |
| | $Cu_2GaPt$ | −0.112 | −0.322 | $Cu + GaCuPt_2 + Ga_3Pt_2$ | −0.015 |
| $Ti_2YZ$ | $Ti_2AuTl$ | −0.064 | −0.109 | $Ti_3Au + TiAu + Tl$ | 0.172 |
| | $Ti_2CdPt$ | −0.082 | −0.367 | $TiPt + Ti_3Pt + Cd$ | 0.191 |
| | $Ti_2PbPd$ | −0.084 | −0.211 | $Ti_3Pb + Ti_2Pd + Pb$ | 0.129 |
| | $Ti_2InAu$ | −0.094 | −0.257 | $Ti_3Au + In + In_2Au$ | 0.037 |

[a] The symbol in the parentheses for the single phase indicates its space-group type.

Table IV. Surviving 133 alloys with the 6M martensitic structure after the materials screening.

| | This study | | | Data from MPD [21] | Difference |
|---|---|---|---|---|---|
| | Alloy | $\Delta E^{m\text{-}p}$ (eV/atom) | $E_f^m$ (eV/atom) | Reported convex hull [a] | $E_f^m - E_f^{\mathrm{conv}}$ (eV/atom) |
| *XY* | MgIn | −0.005 | −0.062 | MgIn (*P*4/*mmm*) | −0.008 |
| | ZrCd | −0.007 | −0.081 | Zr + $ZrCd_3$ | −0.009 |
| | PdAg | −0.029 | −0.032 | Pd + $PdAg_3$ | 0.004 |
| | AgAu | −0.033 | −0.052 | Ag + $AlAu_3$ | −0.031 |
| | AlCu | −0.040 | −0.183 | AlCu (*P*2/*m*) | 0.030 |
| | AlAg | −0.074 | −0.031 | Al + $AlAg_3$ | 0.022 |
| | TiAl | −0.100 | −0.354 | TiAl (*P*4/*mmm*) | 0.049 |
| | InHf | −0.112 | −0.049 | $In_5Hf_2$ + Hf | 0.058 |
| | NiRe | −0.356 | −0.024 | Ni + $NiRe_3$ | 0.050 |
| $X_3Y$ | $Li_3Cd$ | −0.001 | −0.185 | $Li_3Cd$ (*I*4/*mmm*) | −0.007 |
| | $Co_3Ga$ | −0.002 | −0.102 | Co + CoGa | 0.037 |
| | $Li_3Hg$ | −0.002 | −0.287 | $Li_3Hg$ (*Fm*-3*m*) | −0.009 |
| | $Li_3Zn$ | −0.003 | −0.108 | Li + LiZn | −0.002 |
| | $Li_3Ag$ | −0.003 | −0.161 | $Li_3Ag$ (*I*4/*mmm*) | −0.001 |
| | $Na_3In$ | −0.003 | −0.060 | Na + $Na_2In$ | 0.030 |
| | $Na_3Tl$ | −0.003 | −0.089 | $Na_3Tl$ (*Pm*-3*m*) | −0.006 |
| | $Li_3Rh$ | −0.004 | −0.042 | Li + LiRh | 0.056 |
| | $Li_3Au$ | −0.004 | −0.412 | $Li_3Au$ (*Fm*-3*m*) | −0.006 |
| | $Li_3Cu$ | −0.004 | −0.003 | Li + $LiCu_3$ | 0.008 |
| | $Na_3Cd$ | −0.005 | −0.059 | $Na_3Cd$ (*P*6$_3$/*mmc*) | −0.012 |
| | $Cu_3Li$ | −0.016 | −0.040 | $Cu_3Li$ (*I*4/*mmm*) | −0.009 |
| | $Pd_3Hg$ | −0.018 | −0.110 | $Pd_3Hg$ (*Pm*-3*m*) | 0.015 |
| | $Au_3Pd$ | −0.018 | −0.071 | $Au_3Pd$ (*Pm*-3*m*) | 0.009 |
| | $Pd_3Zn$ | −0.019 | −0.315 | $Pd_3Zn$ (I4/mmm) | 0.006 |
| | $Ag_3Na$ | −0.020 | −0.012 | Ag + $Ag_2Na$ | 0.062 |
| | $Ag_3Li$ | −0.020 | −0.147 | $Ag_3Li$ (*I*4/*mmm*) | −0.009 |
| | $Pd_3Pb$ | −0.021 | −0.249 | $Pd_3Pb$ (*Pm*-3*m*) | 0.060 |
| | $Pd_3Bi$ | −0.022 | −0.247 | $Pd_3Bi$ (*Pcmm*) | 0.022 |
| | $Cu_3Al$ | −0.022 | −0.192 | $Cu_3Al$ (*Pmnm*) | −0.002 |
| | $Cu_3Ga$ | −0.022 | −0.113 | $Cu_3Ga$ (*P*6$_3$/*mmc*) | −0.014 |
| | $Pd_3Au$ | −0.023 | −0.025 | $Pd_3Au$ (*Pm*-3*m*) | 0.043 |

| | | | | |
|---|---|---|---|---|
| $Cu_3Be$ | −0.025 | −0.012 | Cu + $CuBe_2$ | 0.042 |
| $Au_3Ag$ | −0.025 | −0.037 | $Au_3Ag$ ($P6_3/mmc$) | −0.005 |
| $Ag_3Mg$ | −0.025 | −0.169 | $Ag_3Mg$ ($P6_3/mmc$) | −0.003 |
| $Ag_3Cd$ | −0.025 | −0.051 | $Ag_3Cd$ ($P6_3/mmc$) | −0.002 |
| $Cu_3Zn$ | −0.025 | −0.072 | Cu + $Cu_5Zn_8$ | −0.031 |
| $Ag_3Pd$ | −0.026 | −0.048 | $Ag_3Pd$ ($I4/mmm$) | 0.006 |
| $Pd_3Cd$ | −0.027 | −0.249 | $Pd_3Cd$ ($I4/mmm$) | 0.011 |
| $Ag_3Zn$ | −0.027 | −0.028 | $Ag_3Zn$ ($Pm$-3$m$) | 0.001 |
| $Cu_3Mg$ | −0.028 | −0.076 | Cu + $Cu_2Mg$ | 0.032 |
| $Pd_3Mg$ | −0.028 | −0.469 | $Pd_3Mg$ ($I4/mmm$) | 0.028 |
| $Ag_3Au$ | −0.029 | −0.040 | Ag + $AgAu_3$ | −0.029 |
| $Au_3Sn$ | −0.029 | −0.072 | Au + AuSn | 0.019 |
| $Cu_3Pd$ | −0.029 | −0.093 | $Cu_3Pd$ ($Pm$-3$m$) | 0.016 |
| $Cu_3Pt$ | −0.030 | −0.121 | $Cu_3Pt$ ($Pm$-3$m$) | 0.017 |
| $Au_3Cd$ | −0.030 | −0.130 | $Au_3Cd$ ($P6_3/mmc$) | −0.002 |
| $Au_3Cu$ | −0.031 | −0.018 | Au + AuCu | 0.005 |
| $Ni_3Be$ | −0.032 | −0.155 | Ni + NiBe | 0.078 |
| $Ni_3Zn$ | −0.032 | −0.109 | Ni + NiZn | 0.016 |
| $Au_3Zn$ | −0.033 | −0.126 | $Au_3Zn$ ($I4_1/acd$) | 0.021 |
| $Cu_3Au$ | −0.034 | −0.040 | $Cu_3Au$ ($Pm$-3$m$) | −0.005 |
| $Sn_3Mg$ | −0.037 | −0.013 | Sn + $SnMg_2$ | 0.059 |
| $Mg_3In$ | −0.037 | −0.061 | $Mg_3In$ ($Pm$-3$m$) | 0.034 |
| $Au_3Mn$ | −0.044 | −0.084 | $Au_4Mn$ + $Au_2Mn$ | 0.006 |
| $Pd_3Cu$ | −0.045 | −0.045 | Pd + $PdCu_3$ | −0.008 |
| $Ag_3Sc$ | −0.046 | −0.227 | $Ag_4Sc$ + AgSc | −0.016 |
| $Pb_3Na$ | −0.047 | −0.093 | $Pb_3Na$ ($Pm$-3$m$) | 0.021 |
| $Cu_3Sc$ | −0.047 | −0.211 | $Cu_3Sc$ ($I4/mmm$) | −0.043 |
| $Pd_3Li$ | −0.051 | −0.230 | $Pd_3Li$ ($Pm$-3$m$) | 0.036 |
| $Ag_3Hf$ | −0.054 | −0.007 | Ag + AgHf | 0.053 |
| $Al_3Li$ | −0.059 | −0.052 | Al + AlLi | 0.038 |
| $Pt_3Cd$ | −0.064 | −0.114 | $Pt_3Cd$ ($Pm$-3$m$) | 0.064 |
| $Pt_3Zn$ | −0.065 | −0.243 | $Pt_3Zn$ ($Pm$-3$m$) | 0.072 |
| $Pt_3Cu$ | −0.079 | −0.047 | $Pt_7Cu$ + PtCu | 0.060 |
| $Ti_3In$ | −0.086 | −0.104 | $Ti_3In$ ($P6_3/mmc$) | 0.029 |
| $Ti_3Sn$ | −0.089 | −0.273 | $Ti_3Sn$ ($P6_3/mmc$) | 0.022 |

| | | | | | |
|---|---|---|---|---|---|
| | $Ni_3Pt$ | −0.092 | −0.047 | $Ni_3Pt$ (*Pm*-3*m*) | 0.025 |
| | $Ti_3Al$ | −0.098 | −0.252 | $Ti_3Al$ ($P6_3/mmc$) | 0.026 |
| | $Ti_3Pb$ | −0.102 | −0.033 | $Ti_3Pb$ ($P6_3/mmc$) | 0.029 |
| | $Tc_3Be$ | −0.103 | −0.016 | $Tc + TcBe_3$ | 0.077 |
| | $Ti_3Hg$ | −0.106 | −0.016 | $Ti_3Hg$ (*Pm*-3*n*) | 0.084 |
| | $Ti_3Zn$ | −0.106 | −0.088 | $Ti + Ti_2Zn$ | 0.025 |
| | $Ti_3Ga$ | −0.112 | −0.299 | $Ti_3Ga$ ($P6_3/mmc$) | 0.021 |
| | $Zr_3Al$ | −0.130 | −0.253 | $Zr_3Al$ (*Pm*-3*m*) | 0.046 |
| | $Ni_3Re$ | −0.139 | −0.004 | $Ni + NiRe_3$ | 0.033 |
| | $La_3Sn$ | −0.151 | −0.434 | $La_3Sn$ (*Pm*-3*m*) | 0.060 |
| | $Hf_3Al$ | −0.182 | −0.207 | $Hf + Hf_3Al_2$ | 0.001 |
| | $Al_3Co$ | −0.185 | −0.271 | $Al_9Co_2 + Al_5Co_2$ | 0.143 |
| | $Y_3Bi$ | −0.199 | −0.351 | $Y + Y_5Bi_3$ | 0.128 |
| | $Sc_3Bi$ | −0.214 | −0.352 | $Sc + Sc_5Bi_3$ | 0.044 |
| $Zn_2YZ$ | $Zn_2LiAg$ | −0.006 | −0.185 | $Zn_3Li + ZnLiAg_2$ | −0.011 |
| | $Zn_2CuPt$ | −0.017 | −0.355 | $Zn_8Cu_5 + Zn_{12}Pt_7 + Cu$ | −0.012 |
| | $Zn_2NiAu$ | −0.020 | −0.203 | $ZnNi + ZnAu$ | 0.023 |
| | $Zn_2MnPt$ | −0.034 | −0.340 | $Zn_3Pt + Zn_{12}Pt_7 + Mn$ | 0.003 |
| | $Zn_2FePt$ | −0.037 | −0.212 | $Zn_3Pt + Zn_{12}Pt_7 + Fe$ | 0.167 |
| | $Zn_2MgPt$ | −0.040 | −0.492 | $Zn_2Mg + Zn_3Pt + Mg_2Pt$ | −0.034 |
| | $Zn_2MgPd$ | −0.049 | −0.417 | $Zn_2Mg + Zn_3Pd + MgPd$ | −0.020 |
| | $Zn_2ScRh$ | −0.086 | −0.569 | $Zn_{11}Rh_2 + Zn_{17}Sc_3 + ScRh$ | −0.002 |
| | $Zn_2AlPt$ | −0.208 | −0.468 | $Zn + Zn_3Pt + Al_3Pt_2$ | 0.076 |
| $Cu_2YZ$ | $Cu_2LiZn$ | −0.009 | −0.120 | $Cu + LiZn$ | −0.014 |
| | $Cu_2PdHg$ | −0.014 | −0.019 | $Cu_3Pd + PdHg + Hg$ | 0.083 |
| | $Cu_2NiPd$ | −0.023 | −0.009 | $Cu_3Pd + Ni + Pd$ | 0.064 |
| | $Cu_2MgAg$ | −0.026 | −0.069 | $Cu + MgAg_3 + Cu_2Mg$ | 0.059 |
| | $Cu_2ZnAg$ | −0.030 | −0.008 | $Cu + Ag + Cu_5Zn_8$ | 0.033 |
| | $Cu_2MgAu$ | −0.032 | −0.242 | $Cu + MgAu$ | 0.063 |
| | $Cu_2NiZn$ | −0.037 | −0.093 | $Cu + NiZn$ | 0.032 |
| | $Cu_2NiPt$ | −0.038 | −0.055 | $Cu_3Pt + Ni_3Pt$ | 0.061 |
| | $Cu_2MnAu$ | −0.040 | −0.017 | $Cu + Mn + MnAu_2$ | 0.023 |
| | $Cu_2ScAu$ | −0.043 | −0.403 | $Cu + ScAu$ | −0.003 |
| | $Cu_2PdIn$ | −0.043 | −0.200 | $Cu + Pd_2In_3 + Pd_5In_3$ | 0.055 |
| | $Cu_2ZnIr$ | −0.045 | −0.009 | $Cu + Ir + CuZn_2Ir$ | 0.065 |
| | $Cu_2ZnRh$ | −0.049 | −0.124 | Cu + ZnRh | 0.069 |
| | $Cu_2MnPd$ | −0.051 | −0.085 | Cu + Mn + $CuMnPd_2$ | 0.027 |
| | $Cu_2ZnPd$ | −0.052 | −0.223 | Cu + ZnPd | 0.062 |
| | $Cu_2ZnPt$ | −0.054 | −0.252 | Cu + ZnPt | 0.032 |
| | $Cu_2NbPd$ | −0.056 | −0.089 | Cu + Nb + $NbPd_3$ | 0.053 |
| | $Cu_2AlGa$ | −0.062 | −0.109 | $CuGa_2$ + $Cu_9Al_4$ + CuAl | 0.056 |
| | $Cu_2AlNi$ | −0.063 | −0.283 | Cu + AlNi | 0.046 |
| | $Cu_2MnPt$ | −0.064 | −0.163 | Cu + MnPt | 0.014 |
| | $Cu_2NiGa$ | −0.065 | −0.183 | Cu + $Ni_2Ga_3$ + $Ni_{13}Ga_9$ | 0.017 |
| | $Cu_2VPd$ | −0.072 | −0.007 | $Cu_3Pd$ + $V_3Pd$ | 0.111 |
| | $Cu_2ZrPd$ | −0.098 | −0.380 | $Cu_5Zr$ + ZrPd + $ZrPd_3$ | −0.015 |
| | $Cu_2ScPt$ | −0.101 | −0.593 | Cu + ScPt | 0.017 |
| | $Cu_2TiPd$ | −0.103 | −0.295 | Cu + $Cu_3Ti$ + $Ti_2Pd_3$ | −0.012 |
| | $Cu_2VPt$ | −0.114 | −0.165 | Cu + VPt | 0.114 |
| | $Cu_2NiNb$ | −0.116 | −0.035 | Cu + Nb + $Ni_3Nb$ | 0.061 |
| | $Cu_2ScIr$ | −0.116 | −0.421 | Cu + ScIr | 0.105 |
| | $Cu_2PdHf$ | −0.119 | −0.403 | Cu + $Pd_3Hf$ + $Cu_8Hf_3$ | −0.006 |
| | $Cu_2NiTa$ | −0.133 | −0.063 | Cu + $NiTa_2$ | 0.081 |
| | $Cu_2ZrRu$ | −0.205 | −0.222 | Cu + CuZr | 0.108 |
| $Ti_2YZ$ | $Ti_2AlCd$ | −0.051 | −0.119 | $Ti_3Al$ + Cd + TiAl | 0.121 |
| | $Ti_2RuHg$ | −0.060 | −0.172 | $Ti_3Hg$ + Hg + TiRu | 0.247 |
| | $Ti_2GaIn$ | −0.076 | −0.185 | $Ti_2Ga$ + In | 0.114 |
| | $Ti_2ZnSn$ | −0.087 | −0.206 | $Ti_5Sn_3$ + $TiZn_3$ | 0.087 |
| | $Ti_2AlIn$ | −0.088 | −0.176 | $Ti_3In$ + TiAl + In | 0.069 |
| | $Ti_2AlGa$ | −0.093 | −0.359 | TiGa + TiAl | 0.076 |
| | $Ti_2AlTl$ | −0.101 | −0.035 | $Ti_3Al$ + Tl + TiAl | 0.205 |
| | $Ti_2MgSn$ | −0.116 | −0.138 | $Ti_2Sn$ + Mg | 0.113 |
| | $Ti_2LiSb$ | −0.120 | −0.244 | $Ti_3Sb$ + $Li_3Sb$ | 0.219 |
| | $Ti_2PbAl$ | −0.130 | −0.016 | TiAl + $Ti_3Al$ + Pb | 0.224 |
| | $Ti_2AlSn$ | −0.158 | −0.258 | $Ti_6Sn_5$ + TiAl + $TiAl_2$ | 0.129 |
| | $Ti_2MgSb$ | −0.180 | −0.178 | $Ti_3Sb$ + $Ti_5Sb_3$ + Mg | 0.163 |
| | $Ti_2LiBi$ | −0.194 | −0.060 | Ti + $Ti_2Bi$ + $Li_3Bi$ | 0.182 |

[a] The symbol in the parentheses for the single phase indicates its space-group type.

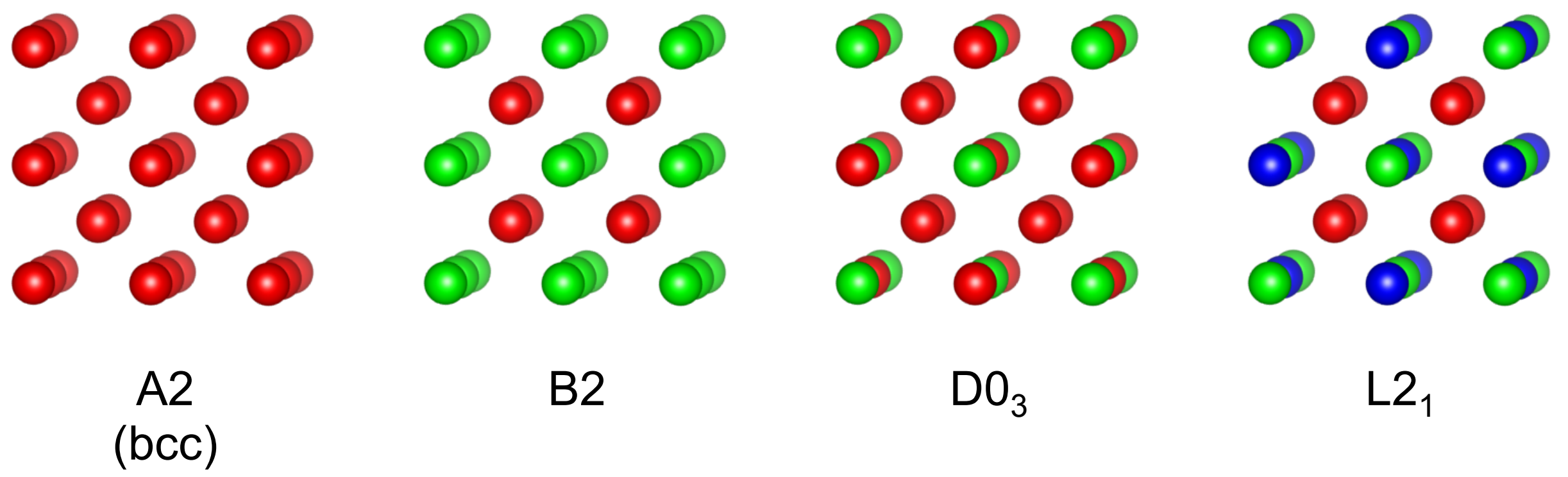

FIG. 1. (Color online) Crystal structures of the parent phase of the investigated alloys as well as the A2 (bcc) crystal structures. Spheres with different colors are for different chemical elements.

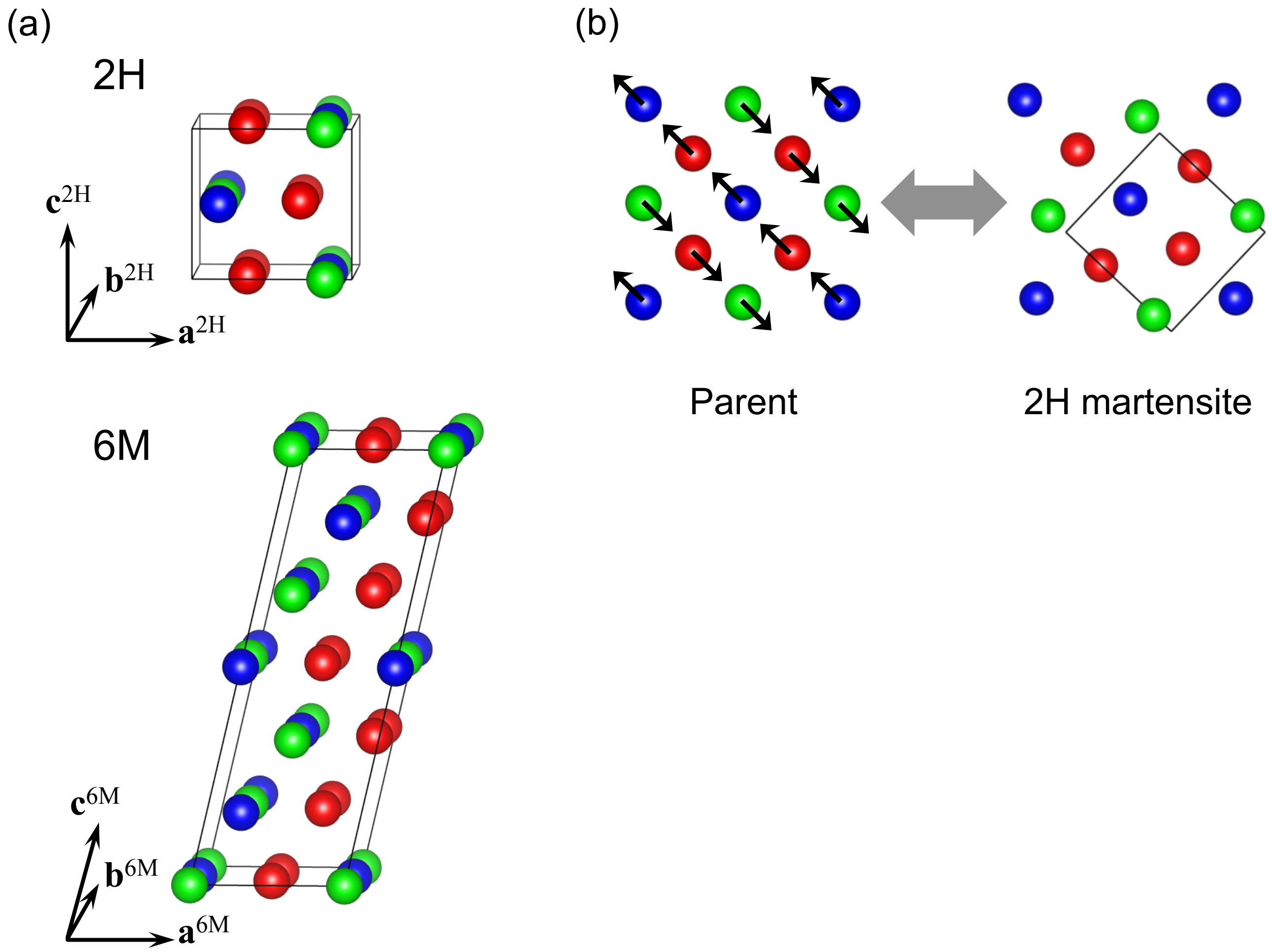

FIG. 2. (Color online) (a) Crystal structure of the 2H and the 6M martensitic phases for the ternary alloys with the L2$_1$ parent phase. (b) Transformation between the parent and the martensitic phases. Small black arrows on spheres denote the atomic movement, and the rectangle denotes the unit cell of the 2H crystal structure.

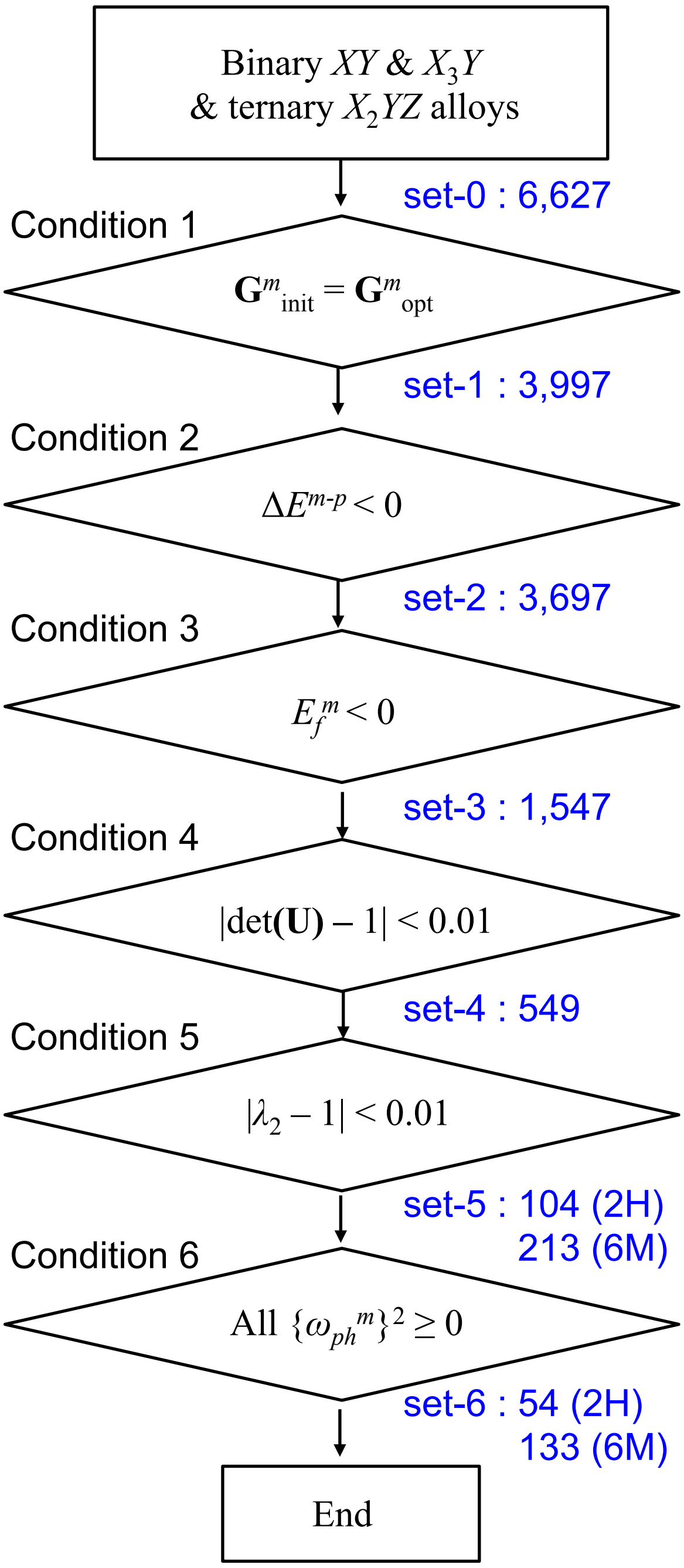

FIG. 3. (Color online) Flowchart of the materials screening. The number of surviving alloys at each step is shown in blue.

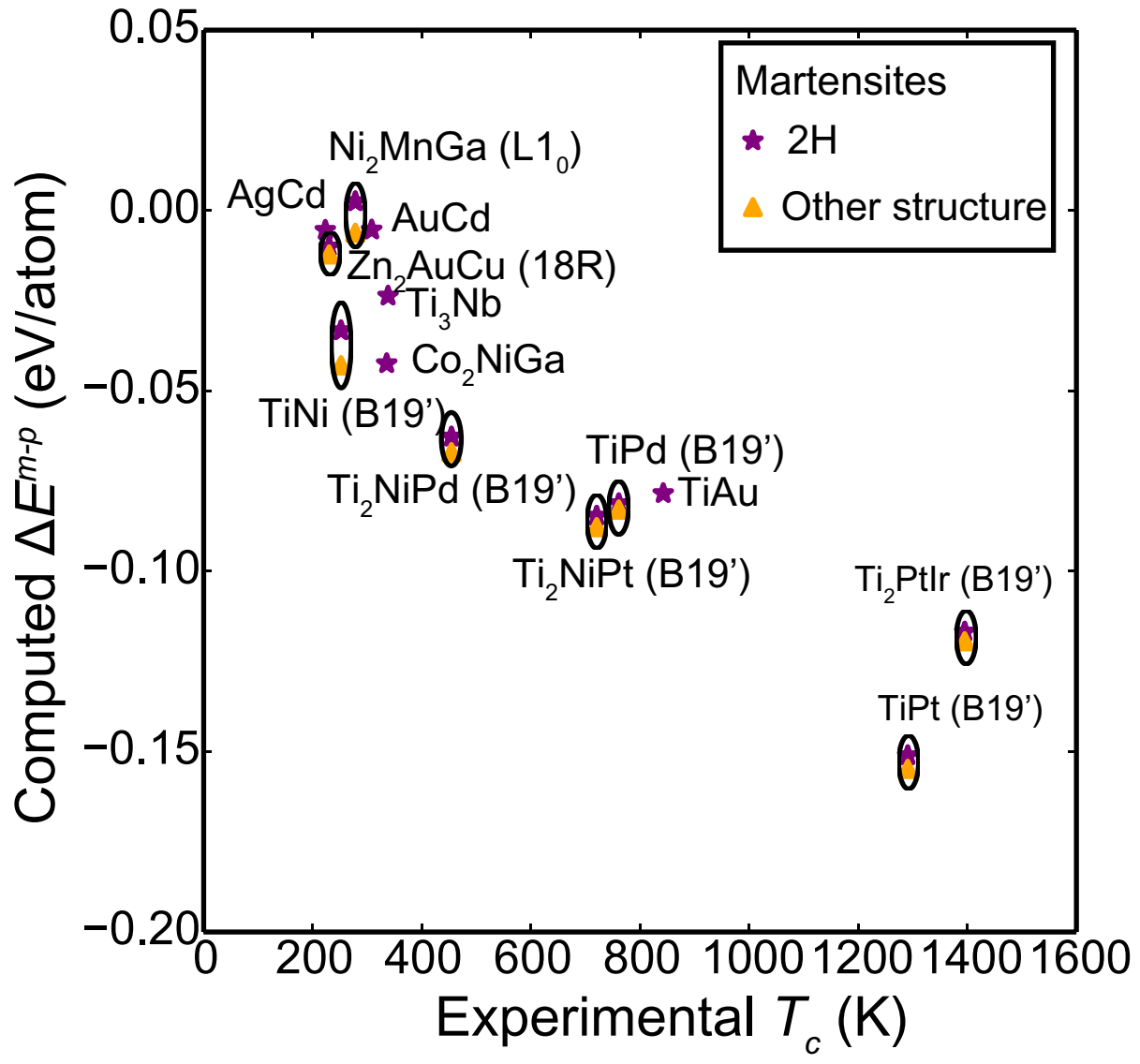

FIG. 4. (Color online) Relationship of computed $\Delta E^{m\text{-}p}$ and experimental transformation temperature $T_c$. The parentheses denote other martensitic phases than the 2H if they are reported in experiments.

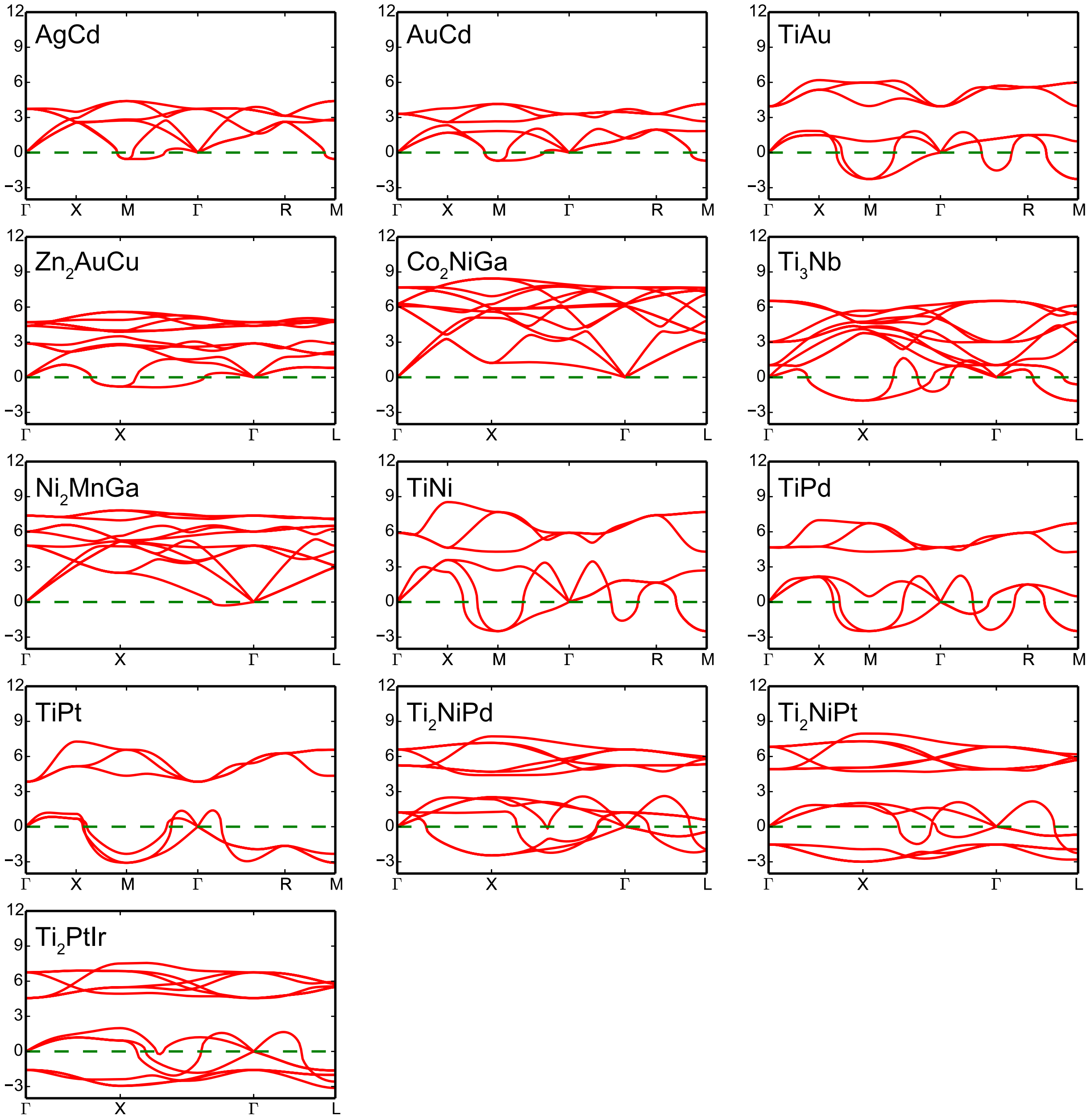

FIG. 5. (Color online) Phonon dispersion curves for the parent phases of the 13 SMAs with stoichiometric composition ratios.

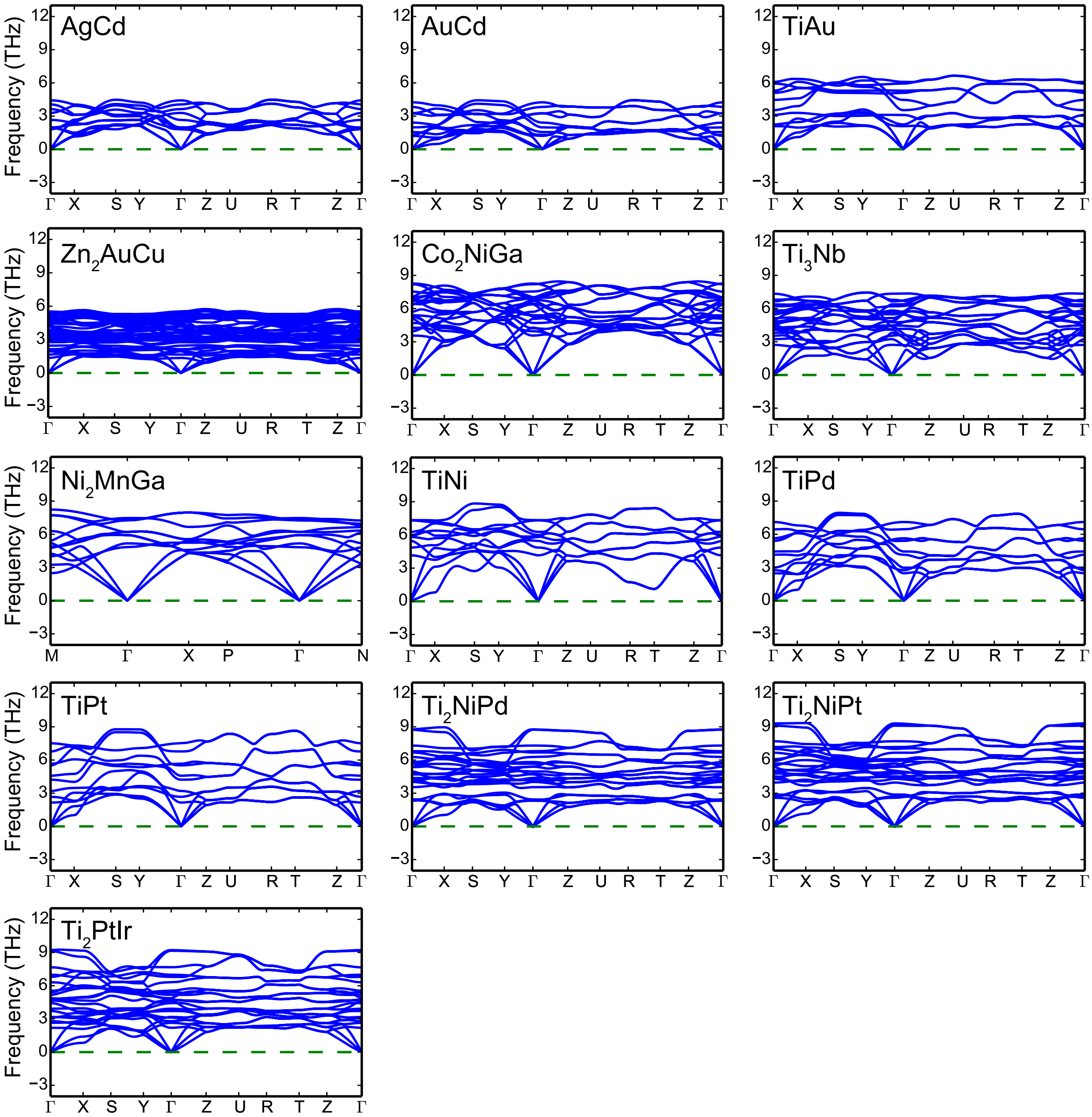

FIG. 6. (Color online) Phonon dispersion curves for the martensitic phases of the 13 SMAs with stoichiometric composition ratios.

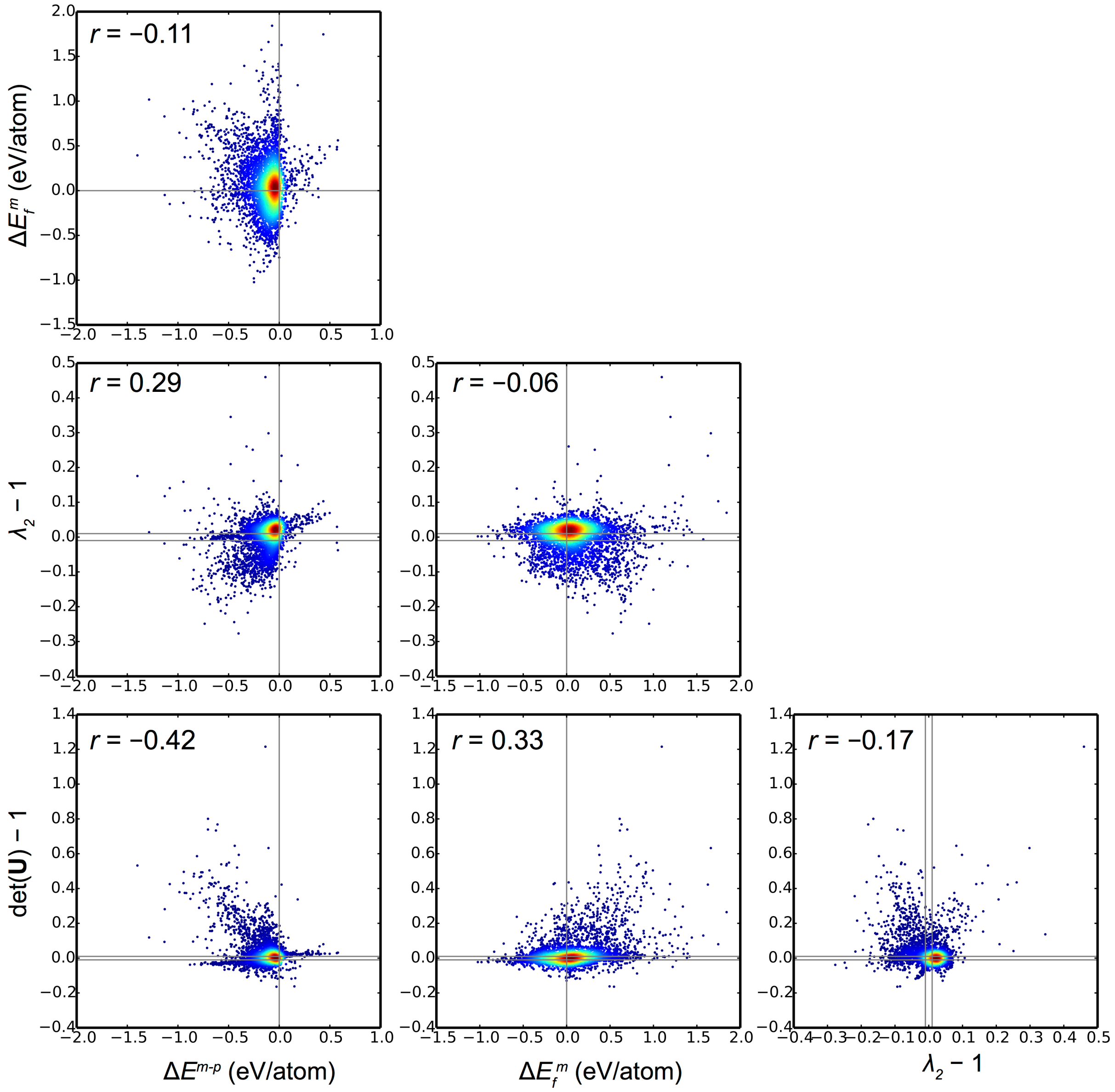

FIG. 7. (Color online) Scatter plots between the quantities associated with the screening conditions for 3,997 alloys in the set-1 in Fig.3. The gray lines denote the boundary of screening-condition slot. The color scale represents the relative density of points generated by Gaussian kernel density estimation. Blue and red colors correspond to higher and lower densities, respectively. $r$ shown in each panel is the linear correlation coefficient.

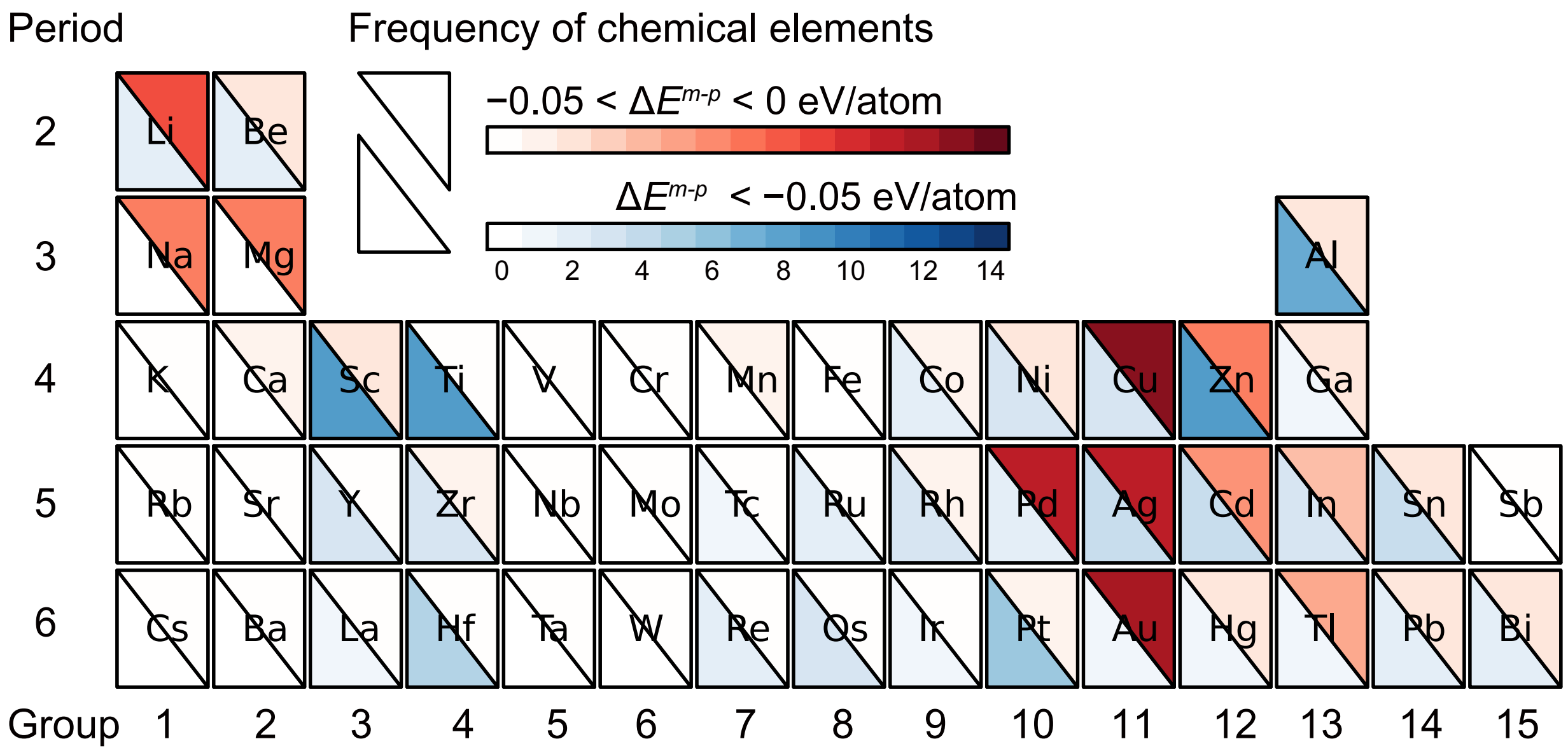

FIG. 8. (Color online) Frequency of chemical elements in the surviving 111 binary alloys. Upward and downward triangles classify $\Delta E^{m\text{-}p}$.